\documentclass[letterpaper, 10 pt, conference]{ieeeconf}  
\usepackage{tikz}
\usepackage{xcolor}
\usepackage{float}
\usepackage{booktabs}
\usepackage{url}
\usepackage{cite}
\usepackage{hyperref}
\definecolor{motifgreen}{RGB}{0,153,0}
\usetikzlibrary{arrows.meta}

\IEEEoverridecommandlockouts                              

\usepackage{graphicx} 
\usepackage{amsmath} 
\newtheorem{theorem}{Theorem}
\usetikzlibrary{arrows.meta,plotmarks}

\newtheorem{corollary}{Corollary}
\newtheorem{lemma}{Lemma}
\title{\LARGE \bf
Feedback and Resource Competition in Incoherent Feedforward Loops: When Is Cumulative Dose Ordering Preserved?}

\author{Iva Janković$^{1}$, Fabiana Calandra$^{2}$,  Giansimone Perrino$^{3}$, Velia Siciliano$^{2}$  and Filippo Menolascina$^{4}$
\thanks{$^{1}$Iva Janković is with the School of Informatics, University of Edinburgh, Edinburgh, UK and supported by the United Kingdom Research and Innovation (grant EP/Y030869/1), UKRI AI Centre for Doctoral Training in Biomedical Innovation. 
\texttt{iva.jankovic@ed.ac.uk}}
\thanks{$^{2}$Fabiana Calandra and Velia Siciliano are with the Center
for Advanced Biomaterials for Healthcare, IIT, Naples, Italy;
Fabiana Calandra is also with University of Naples Federico II.
\texttt{\{fabiana.calandra,velia.siciliano\}@iit.it}}%
\thanks{$^{3}$Giansimone Perrino is an independent researcher. 
        \texttt{ giansimone@gmail.com }}%
\thanks{$^{4}$Filippo Menolascina is with the School of Engineering, University of Edinburgh, Edinburgh, UK. 
        \texttt{ filippo.menolascina@ed.ac.uk}}%
}

\begin{document}

\maketitle
\thispagestyle{empty}
\pagestyle{empty}

\begin{abstract}
Biochemical coupling can change the expression levels and timing
predicted by an incoherent feedforward loop (IFFL) model without
changing cumulative dose ordering. We derive ordering conditions
using shared-production identities and Metzler comparison bounds,
with tail guarantees beyond finite computation. A nonlinear example
exhibits delayed slope reversal at a fixed dose, on an $O(1/\eta)$
timescale as pathway mismatch $\eta$ vanishes. At one fixed joint
fit, validated bounds establish decreasing cumulative output
across 20--100 ng nominal burden, from zero to full target loading,
at every positive finite observation time for a translation-blocking
model and an RNA-decay model. Full loading at 20 ng reduces
48-hour cumulative output by $13.35\%$ and $10.18\%$, respectively,
while endpoint concentration in the first model changes by only
$+0.15\%$. Preserved ordering therefore does not ensure quantitative
agreement.
\end{abstract}

\section{Introduction}
\label{sec:introduction}
\begin{figure}[H]
\centering
\resizebox{0.9\columnwidth}{!}{%
\begin{tikzpicture}[
    font=\scriptsize,
    species/.style={
        circle,draw,minimum size=4.5mm,inner sep=0pt
    },
    direct/.style={
        draw=green!50!black,
        -{Latex[length=1.4mm]},line width=.7pt
    },
    indirect/.style={
        draw=red,
        -{Latex[length=1.4mm]},line width=.7pt
    },
    repression/.style={
        draw=red,
        -{Bar[width=3.5pt]},line width=.7pt
    }
]
\path[use as bounding box] (0,0) rectangle (8.5,2.8);

\node[anchor=west,font=\scriptsize\bfseries]
    at (0,2.6) {(a)};

\node[species] (X) at (.6,1.55) {$X$};
\node[species] (Z) at (1.95,1.55) {$Z$};
\node[species] (Y) at (3.3,1.55) {$Y$};

\node at (.6,2.0) {Input};
\node at (1.95,2.0) {Regulator};
\node at (3.3,2.0) {Output};

\node[text=red] at (1.95,2.5) {Indirect pathway};

\draw[indirect]
    (X) -- node[above,text=red] {$+$} (Z);
\draw[repression]
    (Z) -- node[above,text=red] {$-$} (Y);

\draw[direct,rounded corners=3pt]
    (X.south) -- (.6,.75)
    -- node[below=3pt,text=green!50!black]
       {Direct pathway}
    (3.3,.75) -- node[right,text=green!50!black] {$+$} (Y.south);

\node[anchor=west,font=\scriptsize\bfseries]
    at (4.35,2.6) {(b)};

\draw[-{Latex[length=1.4mm]},line width=.5pt]
    (5.05,.55) -- (8.4,.55);
\draw[-{Latex[length=1.4mm]},line width=.5pt]
    (5.05,.55) -- (5.05,2.35);

\node[rotate=90] at (4.65,1.45) {Gene expression};
\node at (6.75,.15) {DNA abundance};

\draw[black,line width=.6pt]
    plot[smooth,mark=*,mark size=1.2pt]
    coordinates {
        (5.25,1.00)
        (5.98,1.26)
        (6.70,1.53)
        (7.43,1.79)
        (8.15,2.05)
    };

\draw[blue!50!black,line width=.7pt]
    plot[smooth,mark=o,mark size=1.5pt,
         mark options={fill=white}]
    coordinates {
        (5.25,1.00)
        (5.98,1.00)
        (6.70,1.00)
        (7.43,1.00)
        (8.15,1.00)
    };
\node[
    text=black,
    font=\fontsize{5}{5.5}\selectfont,
    rotate=20,
    inner sep=0pt
] at (6.75,1.85) {Without IFFL};

\node[
    text=blue!50!black,
    font=\fontsize{5}{5.5}\selectfont,
    inner sep=0pt
] at (6.95,1.20) {With IFFL};

\end{tikzpicture}%
}
\caption{%
(a) Opposing IFFL pathways: direct (green) and indirect (red). 
(b) Ideal DNA abundance compensation (schematic).}
\label{fig:iffl_principle}
\end{figure}
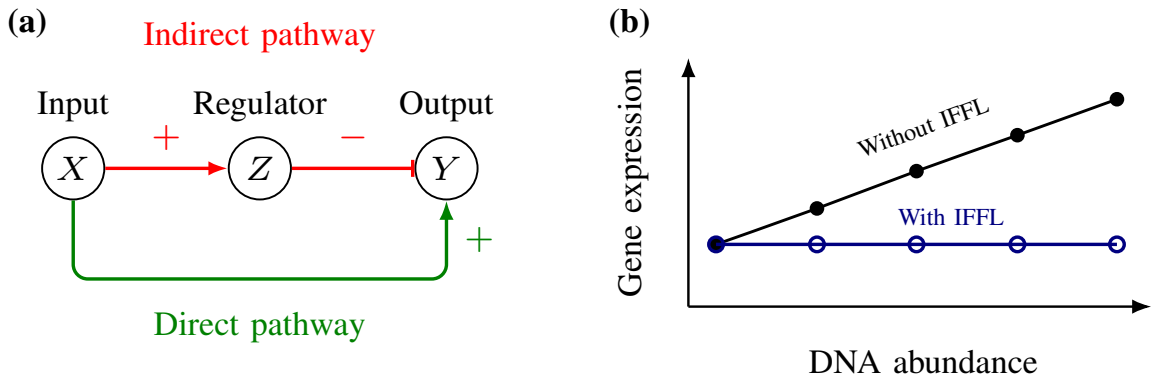

Regulatory motifs help explain how small molecular circuits generate
characteristic biological responses. Depending on their structure and kinetics, these circuits can
generate pulses, filter transient signals or adapt to sustained
inputs \cite{alon2006,mangan2003,ma2009}.  One such circuit is the incoherent feedforward loop (IFFL), in which an input regulates an output through two pathways with opposing effects (Fig.~\ref{fig:iffl_principle}a). The direct pathway promotes output expression, while the indirect pathway increases production of the regulator, which represses the output. These opposing effects can partly offset each other, making the output less sensitive to changes in input.

This principle motivates circuits that reduce the sensitivity
of gene expression to DNA template abundance
(Fig.~\ref{fig:iffl_principle}b), as demonstrated in mammalian
cells \cite{bleris2011}. We study two models of post-transcriptional
repression in which a repressor protein binds the output's
messenger RNA (mRNA), the template used to produce proteins.
Repression blocks translation in one model and promotes mRNA
degradation in the other. Additional coupling can alter the
balance between the pathways, so circuit design requires knowing
when models must include it for reliable predictions.

Target binding and competition for shared resources are two
sources of this additional coupling. The binding that allows
the repressor to inhibit output expression can also change its
availability and dynamics. This effect on the regulator is known
as retroactivity \cite{delvecchio2008,gyorgy2014}. Competition
for the machinery used to produce proteins provides another
source of coupling: increasing demand from one gene can reduce
expression of another \cite{qian2017}. Moriya et al.\ showed
that changing a reporter promoter can alter oscillator dynamics
and used simulations to distinguish regulator binding effects
from competition for shared protein degradation machinery
\cite{moriya2019}.
Related IFFL studies show that loading can change response
time and pulse amplitude \cite{wang2020}. In mammalian cells,
Frei et al.\ characterized resource redistribution and
demonstrated miRNA-based burden compensation using 48-hour
expression measurements \cite{frei2020}; Jones et al.\
developed an endoribonuclease-based controller that adapts
expression to resource loading and examined its temporal
behavior \cite{jones2020}. These findings motivate two
comparisons: how much coupling changes expression levels and
timing, and whether it preserves how cumulative output changes
with dose.

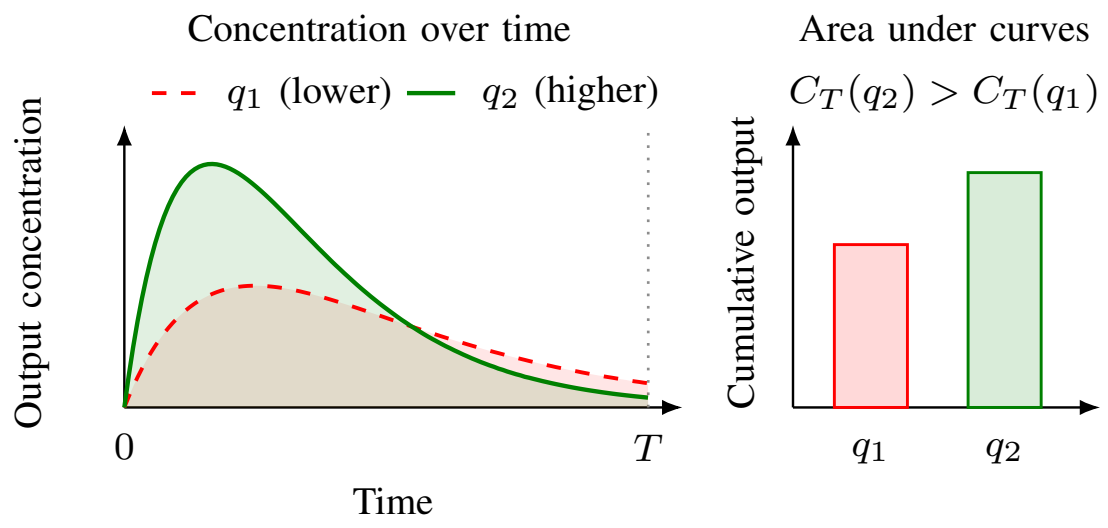
\begin{figure}[H]
\centering
\resizebox{0.9\columnwidth}{!}{%
\begin{tikzpicture}[x=1cm,y=1cm,
 font=\fontsize{8}{9}\selectfont,
 every node/.style={align=center},
 doseone/.style={red,dashed,line width=.8pt},
 dosetwo/.style={green!50!black,line width=.9pt},
 axis/.style={-{Latex[length=1.5mm]},line width=.5pt}]
\path[use as bounding box] (0,0) rectangle (8.4,3.82);
\node at (2.95,3.60) {Concentration over time};
\node at (7.05,3.60) {Area under curves};

\draw[doseone] (1.30,3.13) -- (1.65,3.13);
\node[anchor=west] at (1.72,3.13) {$q_1$ (lower)};
\draw[dosetwo] (3.15,3.13) -- (3.50,3.13);
\node[anchor=west] at (3.57,3.13) {$q_2$ (higher)};

\draw[axis] (1.10,.85) -- (5.15,.85);
\draw[axis] (1.10,.85) -- (1.10,2.90);
\node[rotate=90] at (.43,1.90) {Output concentration};
\node at (3.05,.17) {Time};
\node[below=2pt] at (1.10,.85) {$0$};
\draw[black!45,dotted,line width=.55pt] (4.90,.85) -- (4.90,2.88);
\node[below=2pt] at (4.90,.85) {$T$};

\begin{scope}[shift={(1.10,.85)},x=.95cm,y=2.4cm]
 \fill[red,opacity=.10]
  (0,0) -- plot[domain=0:4,samples=120] (\x,{\x*exp(-\x)})
  -- (4,0) -- cycle;
 \fill[green!50!black,opacity=.12]
  (0,0) -- plot[domain=0:4,samples=120] (\x,{3*\x*exp(-1.5*\x)})
  -- (4,0) -- cycle;
 \draw[doseone,domain=0:4,samples=120]
  plot (\x,{\x*exp(-\x)});
 \draw[dosetwo,domain=0:4,samples=120]
  plot (\x,{3*\x*exp(-1.5*\x)});
\end{scope}

\draw[axis] (5.95,.85) -- (8.18,.85);
\draw[axis] (5.95,.85) -- (5.95,2.90);
\node[rotate=90] at (5.60,1.90) {Cumulative output};
\pgfmathsetmacro{\Clow}{1-5*exp(-4)}
\pgfmathsetmacro{\Chigh}{4/3*(1-7*exp(-6))}
\filldraw[fill=red!15,draw=red,line width=.6pt]
 (6.25,.85) rectangle (6.78,{.85+1.30*\Clow});
\filldraw[fill=green!50!black!15,draw=green!50!black,line width=.6pt]
 (7.22,.85) rectangle (7.75,{.85+1.30*\Chigh});
\node[below=3pt] at (6.515,.85) {$q_1$};
\node[below=3pt] at (7.485,.85) {$q_2$};
\node at (7.05,3.13) {$C_T(q_2)>C_T(q_1)$};
\end{tikzpicture}
}
\caption{Endpoint and cumulative outputs for $q_2>q_1$ (schematic).}
\label{fig:cdr_ordering}
\end{figure}

Cumulative output is the area under the output concentration
curve over a common observation period. We test whether
coupling preserves the reference model's increasing or
decreasing dose dependence. In Fig.~\ref{fig:cdr_ordering},
the higher dose gives a lower endpoint concentration but
a larger cumulative output. This distinction matters for
IFFLs whose transient responses can differ despite
equal steady-state outputs.

Gupta and Sontag proved monotone cumulative responses for two
canonical IFFLs, even when instantaneous dose responses can be
nonmonotone \cite{gupta2025}. Wafi, de Oliveira and Sontag
developed integral-kernel sign tests for a larger class with
nonlinear output dynamics and a linear regulator subsystem
independent of downstream states \cite{wafi2026}. The repression
models studied here lack this structure: binding and shared
translation make regulator dynamics nonlinear and dependent on
downstream target RNA. Increasing burden can reduce output
translation while also reducing repressor production and
relieving repression. The integral formula and its conditional
sign arguments remain valid, but the required signs must be
established along these coupled trajectories.

We separate competing sensitivity contributions using
shared-production and molecular-balance identities, then bound
them with Metzler comparison systems.
Section~\ref{sec:preliminaries} defines the models and response
measures. Section~\ref{sec:structural} derives ordering conditions,
delayed-reversal scaling and tail bounds.
Section~\ref{sec:repression} compares the two repression models and
certifies cumulative ordering at one fixed fit.
Section~\ref{sec:discussion} discusses implications and limitations.

\section{Models and Measures}
\label{sec:preliminaries}
Let $q\in I=[q_-,q_+]$ denote dose and $\alpha\in[0,1]^m$
normalized coupling strengths: $\alpha_e=0$ omits coupling $e$,
while $\alpha_e=1$ gives its prescribed full strength.
The state $X$ and measured output concentration $Y$ satisfy
\[
 \dot X=f(t,X,q,\alpha),\quad X(0)=X_0(q),\quad Y=e_Y^\top X,
\]
where $e_Y$ selects the output.\footnote{
Initial conditions and all other kinetic parameters are fixed across
models at each dose. Assume $f$ is continuous in time with jointly
continuous derivatives through order two in $(X,q,\alpha)$, and
$X_0$ is $C^2$. On each finite horizon, solutions exist and remain in a
common compact subset of an open region where these
conditions hold. } Cumulative dose ordering concerns the area under the output curve,
viewed as a function of dose \cite{gupta2025,wafi2026}. For $T>0$, define
\[
 C_T=\int_0^T Y(t;q,\alpha)\,\mathrm{d}t,\quad
 S_T=\partial_q C_T.
\]
At fixed $T$ and $\alpha$, let $\sigma\in\{-1,1\}$
specify the desired direction. The condition $\sigma S_T>0$
throughout $I$ guarantees strictly increasing ($\sigma=1$)
or strictly decreasing ($\sigma=-1$) cumulative output with dose.

We distinguish changes in expression from preservation of this ordering.
For the reference values $C_T^0=C_T(q,0)$ and $S_T^0=S_T(q,0)$,
define\footnote{The normalized measures assume nonnegative output,
$C_T^0>0$ and a reference slope with a strict, constant sign throughout $I$.}
\[
 E_T=\frac{C_T-C_T^0}{C_T^0},\quad
 R_T=1-\frac{S_T}{S_T^0}.
\]
Here $E_T$ measures relative cumulative-output change.
Since $S_T=(1-R_T)S_T^0$, $0<R_T<1$ means attenuation
and $R_T<0$ amplification, both with the reference sign.
The slope is unchanged at $R_T=0$, zero at $R_T=1$,
and reversed for $R_T>1$. Thus $R_T<1$ preserves sign.
To compare the entire expression trajectory, use
\[
 D_T^\infty=
 \frac{\max_{0\leq t\leq T}|Y(t;q,\alpha)-Y(t;q,0)|}
 {\max_{0\leq t\leq T}Y(t;q,0)}.
\]
For chosen nonnegative tolerances $\tau_Y,\tau_C$, omission
meets the accuracy requirements when $D_T^\infty\leq\tau_Y$
and $|E_T|\leq\tau_C$ throughout the required domain.
Preserving ordering alone guarantees neither accuracy bound.

A positive kernel converts production sensitivity into cumulative
sensitivity. For $\dot Y=V-\delta Y$, with constant $\delta>0$
and dose-independent $Y(0)$, the integral solution gives
\cite{wafi2026}
\begin{equation}
 \lambda_T(t)=\frac{1-e^{-\delta(T-t)}}{\delta},\quad
 S_T=\int_0^T\lambda_T(t)\partial_qV(t)\,\mathrm{d}t.
 \label{eq:kernel}
\end{equation}
The derivative includes dose effects through all internal states.
We quantify trajectory and cumulative-output changes numerically.
The following results give sufficient conditions for preserving
cumulative dose ordering; tail bounds address observation times
beyond the computed interval.
We write $[r]_+=\max\{r,0\}$.

\section{Ordering and Reversal}
\label{sec:structural}

\subsection{Shared production and branch matching}

A common production factor lets us separate branch matching from
the terms that can oppose ordering. Let $x$ denote a regulator,
$z$ the remaining internal states, and $Y$ the output.
With $w=(x,z)$, consider
\begin{align}
 \dot x&=r-\ell+\sum_e\alpha_ej_e,\label{eq:storagex}\\
 \dot Y&=rh-\delta Y+\sum_e\alpha_ev_e.\label{eq:storagey}
\end{align}
Here $\delta>0$ is constant,
$r=r(t,w,q,\alpha)$ is the shared production factor,
$\ell=\ell(t,w)$ is regulator removal,
$h=h(t,w)>0$ modulates output production,
and $j_e=j_e(t,w,q)$, $v_e=v_e(t,w,q)$
describe the additional coupling terms.
Positive constant branch factors are absorbed into $r,h$.
The displayed kinetic functions have the standing state and parameter
regularity, and $h$ is also $C^1$ in time.
Initial values $x(0),Y(0)$ are dose independent;
other initial states may depend on dose.
Biochemical production rates must remain nonnegative.

Let $p=\partial_qx$ denote the regulator's dose sensitivity.
Define the branch mismatch by $\xi_e=v_e-hj_e$, and set
$K_q=\dot x\,\partial_qh-\dot h\,p$.
All dose derivatives include dependence through the internal
trajectories, and $\dot h$ is the total time derivative.
The term $K_q$ accounts for modulation through other states
or explicit time dependence.

\begin{lemma}[Sensitivity identity for shared production]
\label{thm:storage}
For \eqref{eq:storagex}--\eqref{eq:storagey} under these assumptions,
\begin{equation}
 S_T=\mathcal A_T+\int_0^T\lambda_T
 \left(K_q+\sum_e\alpha_e\partial_q\xi_e\right)\mathrm{d}t,
 \label{eq:identity}
\end{equation}
where
\[
 \mathcal A_T=\int_0^T\left[e^{-\delta(T-t)}hp+
 \lambda_T\partial_q(\ell h)\right]\mathrm{d}t.
\]
Suppose uniformly over the dose and coupling ranges that
$\sigma\mathcal A_T\geq A_T>0$, $\sigma K_q\geq-b_0$ and
$\sigma\partial_q\xi_e\geq-b_e$, with nonnegative integrable $b_i$.
Then
\[
 \sigma S_T\geq A_T-B_{0,T}-\sum_e\alpha_eB_{e,T},
\]
with $B_{i,T}=\int_0^T\lambda_Tb_i\,\mathrm{d}t$.
\end{lemma}

\begin{proof}
Eliminate $r$ to obtain
$V=h(\dot x+\ell)+\sum_e\alpha_e\xi_e$.
Differentiation gives
\[
 \partial_qV=\frac{\mathrm{d}}{\mathrm{d}t}(hp)
 +\partial_q(\ell h)+K_q+\sum_e\alpha_e\partial_q\xi_e.
\]
Integrate using \eqref{eq:kernel} and integrate the first term by
parts. Its boundary terms vanish since $p(0)=0$ and
$\lambda_T(T)=0$; $-\dot\lambda_T=e^{-\delta(T-t)}$ gives
\eqref{eq:identity}. Nonnegative weights give the bound.
\end{proof}

Every term follows the coupled trajectory; $\mathcal A_T$ is
generally not the reference slope. For $h=h(x)$ and $\ell=d(x)$,
$K_q=0$ and $\partial_q(\ell h)=(dh)'p$.
Matched contributions $v_e=hj_e$ preserve the desired sign if
$(dh)'\geq0$ and $\sigma p\geq0$, strictly when $\sigma p>0$
on a set of positive measure.
Internal order must hold after coupling.

\subsection{Finite-horizon coupling bounds}
\label{sec:comparison}

A Metzler matrix has nonnegative off-diagonal entries and
supports positive-system comparison \cite{angeli}.
Finite-horizon bounds do not require a Hurwitz matrix
(all eigenvalues with negative real parts); stability is needed
for the infinite-tail bound below.

Let $u\in[u_-,u_+]\subset(0,\infty)$ be a maintained input,
taking the role of dose $q$. Consider the dimensionless model
\begin{align}
 \dot x&=u+\sum_e\alpha_ea_e\phi_e(z_e)-x,\label{eq:returnx}\\
 \dot z_e&=\kappa_e(x-z_e),\label{eq:returnz}\\
 \dot y&=\frac{u+\sum_e\alpha_eb_e\phi_e(z_e)}{x}-y.
 \label{eq:returny}
\end{align}
Assume $0\leq b_e\leq a_e$, $\kappa_e>0$,
$0\leq\phi_e\leq M_e$, $0\leq\phi'_e\leq L_e$ and standing
smoothness. Initial values $x_0>0$, $z_{e0},y_0\geq0$ are fixed.
At $\alpha=0$, the $(x,y)$ subsystem is a canonical IFFL
model \cite{wafi2026}.
Each $z_e$ tracks $x$ and feeds back with branch gains $a_e,b_e$.
Set $x_L=u_-+(x_0-u_-)e^{-t}$,
$\overline x=\max\{x_0,u_++\sum_ea_eM_e\}$ and
$A_T=[1-(1+T)e^{-T}]/\overline x$.
Let $\bar p,\bar s_e$ solve
$\dot{\bar p}=1-\bar p+\sum_ea_eL_e\bar s_e$ and
$\dot{\bar s}_e=\kappa_e(\bar p-\bar s_e)$ from zero.

\begin{theorem}[Ordering under nonlinear feedback]
\label{cor:returnbudget}
For \eqref{eq:returnx}--\eqref{eq:returny}, uniformly over the
stated dose and coupling ranges,
\begin{equation}
 \partial_u C_T\geq A_T-\sum_e\alpha_eB_{e,T},
 \label{eq:activationbound}
\end{equation}
where $B_{e,T}=(a_e-b_e)L_e\int_0^T
(1-e^{-(T-t)})\bar s_e/x_L\,\mathrm{d}t$.
For $\underline x_T=\min\{x_0,x_L(T)\}$ and
$\gamma=\max\{0,\sum_ea_eL_e-1\}$,
\begin{equation}
 B_{e,T}\leq
 \frac{(a_e-b_e)L_e\kappa_e e^{\gamma T}T^4}
 {24\underline x_T}.
 \label{eq:coarse}
\end{equation}
Thus $B_{e,T}=O(T^4)$ whereas
$A_T\sim T^2/(2\overline x)$ as $T$ tends to zero.
\end{theorem}

\begin{proof}
State and sensitivity comparison give $x_L\leq x\leq\overline x$,
$1-e^{-t}\leq p\leq\bar p$ and $0\leq s_e\leq\bar s_e$,
where $p=\partial_ux$ and $s_e=\partial_uz_e$.
Apply Lemma~\ref{thm:storage} with $h=1/x$, $\ell=x$
and $\xi_e=-(a_e-b_e)\phi_e(z_e)/x$.
Since $K_u=\partial_u(\ell h)=0$ and
\[
\begin{aligned}
\partial_u\xi_e
 &=-\frac{(a_e-b_e)\phi'_e(z_e)s_e}{x}
   +\frac{(a_e-b_e)\phi_e(z_e)p}{x^2}\\
 &\geq-\frac{(a_e-b_e)L_e\bar s_e}{x_L},
\end{aligned}
\]
the identity gives \eqref{eq:activationbound}, with positive
contribution at least $A_T$.
The comparison matrix has maximum row sum at most $\gamma$,
so $\bar p\leq te^{\gamma t}$ and
$\bar s_e\leq\kappa_e t^2e^{\gamma t}/2$.
Using $1-e^{-(T-t)}\leq T-t$, $x_L\geq\underline x_T$ and
$\int_0^T(T-t)t^2\,\mathrm{d}t=T^4/12$ gives \eqref{eq:coarse}.
\end{proof}

Matching $a_e=b_e$ eliminates every mismatch budget and preserves
strictly increasing cumulative output at every finite horizon.
For $\phi(z)=z^2/(1+z^2)$, $a=2$, $b=0$, $\kappa=0.05$,
$u\in[0.4,2]$ and $x_0=z_0=y_0=1$, \eqref{eq:coarse} gives
$\partial_uC_1>0.060$ uniformly in dose and coupling despite
an unstable comparison matrix.

\subsection{Observation horizon and loss of ordering}
\label{sec:horizon}

An integrable adverse tail permits a tail test.
Write $Q_\sigma=\sigma\partial_qY$ and
$\omega_\sigma=\sigma\partial_qV$ along a global solution.
Solving $\dot Q_\sigma=\omega_\sigma-\delta Q_\sigma$
from $T_0$ and integrating gives, for every $T\geq T_0$,
\begin{equation}
 \sigma S_T\geq\sigma S_{T_0}-\mathcal B_{\rm tail}(T_0),
 \label{eq:generaltail}
\end{equation}
where $\mathcal B_{\rm tail}(T_0)=[-Q_\sigma(T_0)]_+/\delta
+\delta^{-1}\int_{T_0}^{\infty}[-\omega_\sigma]_+\,\mathrm{d}t$
is assumed finite.
The bound discards positive contributions and bounds the remaining
kernels by $1/\delta$.
An accumulated margin exceeding this budget excludes later loss.
A guarantee over a dose--coupling domain requires uniform
bounds on both the accumulated margin and the adverse tail.

Persistent mismatch behaves differently. In the model with one feedback branch,
take $a=\epsilon$, $b=(1-\eta)\epsilon$, $\alpha=1$ and
$\phi(z)=z^2/(1+z^2)$, with $0<\epsilon<1$, $0<\eta\leq1$.
Its equilibrium satisfies $x=z=\chi$ and
$u=\chi-\epsilon\phi(\chi)$.

\begin{theorem}[First loss under weak mismatch]
\label{cor:firstloss}
Fix $\epsilon\in(0,1)$, $\kappa>0$, a dose with $0<\chi<1$,
and dose-independent $x_0>0$, $z_0,y_0\geq0$.
For sufficiently small $\eta>0$, $\partial_uC_T$ has a unique
positive zero, the first loss of its initial positive sign, with
\begin{equation}
 T_{\rm loss}(\eta)=
 \frac{(1+\chi^2)^2}{\epsilon\eta\chi(1-\chi^2)}+O(1).
 \label{eq:losslaw}
\end{equation}
Here $\eta$ tends to zero with the other parameters fixed.
\end{theorem}

\begin{proof}
Put $p=\partial_ux$ and $s=\partial_uz$.
Lemma~\ref{thm:storage} gives $S_T=A(T)-\epsilon\eta J(T)$, where
\[
\begin{aligned}
A(T)&=\int_0^Te^{-(T-t)}p/x\,\mathrm dt>0,\\
J(T)&=\int_0^T(1-e^{-(T-t)})
\left(\phi'(z)s/x-\phi(z)p/x^2\right)\mathrm dt.
\end{aligned}
\]
Both are independent of $\eta$. Near zero,
$A(T)\sim T^2/(2x_0)$ and $J(T)=O(T^3)$.
Since $\epsilon\max\phi'=9\epsilon/(8\sqrt3)<1$,
states and sensitivities converge exponentially.
Writing $\chi_u=[1-\epsilon\phi'(\chi)]^{-1}$ gives
$A=a_\infty+O(e^{-\lambda T})$, $A'=O(e^{-\lambda T})$,
$J=g_\infty T+O(1)$ and $J'=g_\infty+O(e^{-\lambda T})$,
where $a_\infty=\chi_u/\chi$ and
$g_\infty=\chi_u(1-\chi^2)/(1+\chi^2)^2>0$.
The ratio $J/A$ is bounded on every fixed interval,
including its limit at zero, so small $\eta$ preserves positivity
there. Eventually $A\geq a_\infty/2$ and
$J\leq g_\infty T+C$, so every zero lies beyond $c/\eta$
for some $c>0$. There $S_T'=A'-\epsilon\eta J'<0$ for
sufficiently small $\eta$. Since $S_T\to-\infty$, exactly one
zero exists. At that zero $A=a_\infty+o(\eta)$, yielding
$T=a_\infty/(\epsilon\eta g_\infty)+O(1)$ and \eqref{eq:losslaw}.
\end{proof}

Stable dynamics thus allow reversal of the local slope at a
fixed dose as the horizon changes; this does not establish a
nonmonotone dose-response curve at fixed time. The law is
nonuniform near $\chi=1$ or $\kappa=0$. Delivered RNA decay
creates the different tail regime studied next.

\section{Resource Coupling in Repression Models}
\label{sec:repression}

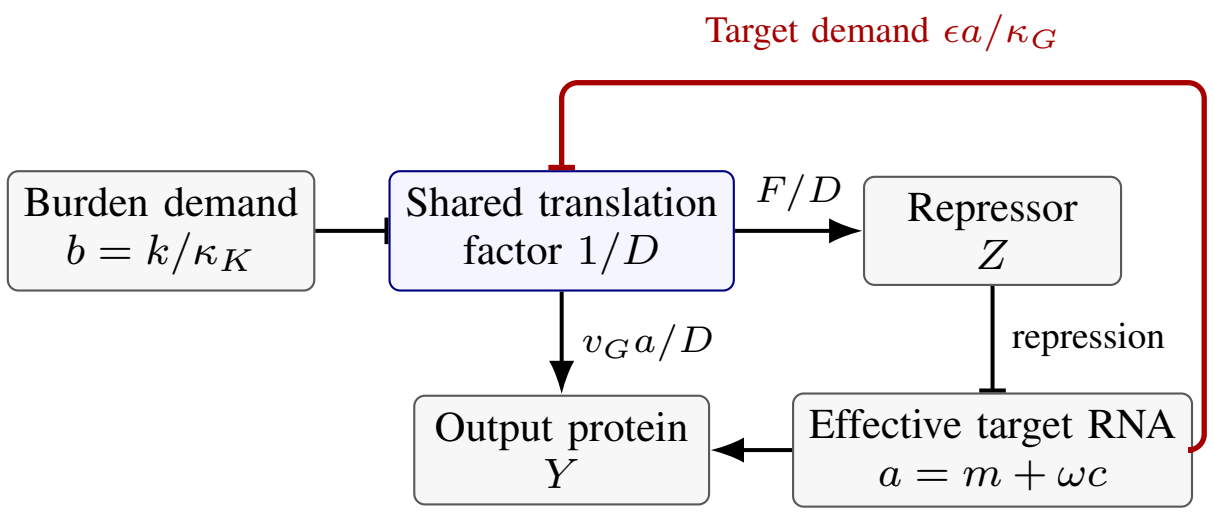
\begin{figure}[H]
\centering
\begin{tikzpicture}[
  x=1cm,y=1cm,>=Latex,
  every node/.style={font=\footnotesize,align=center},
  species/.style={
    draw=black!65,rounded corners=2pt,fill=black!3,
    minimum height=.70cm,inner sep=3pt
  },
  factor/.style={
    species,draw=blue!50!black,fill=blue!4,
    minimum width=1.95cm
  },
  positive/.style={->,line width=.75pt},
  negative/.style={-{Bar[width=5pt]},line width=.75pt},
  demand/.style={
    -{Bar[width=5pt]},line width=1pt,
    draw=red!65!black,rounded corners=3pt
  }
]
  \node[species,minimum width=1.65cm] (b)
    at (-2.65,0) {Burden demand\\$b=k/\kappa_K$};

  \node[factor] (R)
    at (0,0) {Shared translation\\factor $1/D$};

  \node[species,minimum width=1.70cm] (Z)
    at (2.85,0) {Repressor\\$Z$};

  \node[species,minimum width=2.25cm] (a)
    at (2.85,-1.45) {Effective target RNA\\$a=m+\omega c$};

  \node[species,minimum width=1.95cm] (Y)
    at (0,-1.45) {Output protein\\$Y$};

  \draw[negative] (b.east) -- (R.west);

  \draw[positive] (R.east) --
    node[above,font=\scriptsize] {$F/D$}
    (Z.west);

  \draw[positive] (R.south) --
    node[right,font=\scriptsize] {$v_Ga/D$}
    (Y.north);

  \draw[negative] (Z.south) --
    node[right,font=\scriptsize] {repression}
    (a.north);

  \draw[positive] (a.west) -- (Y.east);

  \draw[demand]
    (a.east) -- (4.25,-1.45) -- (4.25,.98)
    -- node[above=2pt,font=\scriptsize,text=red!65!black]
       {Target demand $\epsilon a/\kappa_G$}
    (0,.98) -- (R.north);
\end{tikzpicture}
\caption{Shared translation couples output and repressor production.}
\label{fig:topology}
\end{figure}

Target RNA can reduce the shared translation resources
available to produce its repressor \cite{frei2020,jones2020},
relieving repression and increasing target demand further
(Fig.~\ref{fig:topology}). We consider two mechanisms:
Model~I blocks translation of bound RNA, while Model~II
accelerates its degradation and allows residual translation.
For each mechanism, both compared loading configurations retain
binding; only target translation demand changes.

\subsection{Molecular balances and sensitivity bounds}

Both models retain repressor binding and competition for translation
resources. Let $m,c,Z,Y$ denote free target RNA, bound target RNA,
free repressor and output protein. The totals are $M=m+c$ and $W=Z+c$.
The translation-equivalent target amount is
$a=m+\omega c=M-sc$, where $s=1-\omega$ and $\omega$ is the
relative contribution of bound RNA. Additional bound-RNA decay at
rate $\Delta$ releases repressor. Model~I has $\Delta=\omega=0$;
Model~II uses fixed positive values.
Delivered regulator and burden RNA decay as
$r_R=r_0e^{-\beta t}$ and $k=k_0e^{-\beta t}$.
Define
\[
 b=k/\kappa_K,\quad B=1+r_R/\kappa_R+b,\quad
 D=B+\epsilon a/\kappa_G.
\]
For this single coupling, write $\alpha=\epsilon\in[0,1]$.
The loading strength scales target resource demand in every
translation branch. Its zero value removes this demand from
the model, while retaining repression.
With $e=\eta_+$, $h_b=\beta+\Delta+\eta_-$,
$F=\gamma_RR_{\rm tot}r_R/\kappa_R$ and
$v_G=\gamma_GR_{\rm tot}/\kappa_G$, the balances are
\begin{equation}
 \dot M=-\beta M-\Delta c,\quad \dot c=eZ(M-c)-h_bc,
 \label{eq:repressionbinding}
\end{equation}
\begin{equation}
 \dot W=F/D-dZ,\quad \dot Y=V-dY,\quad V=v_Ga/D.
 \label{eq:repression}
\end{equation}
Assume $\Delta\geq0$, $0\leq\omega\leq1$ and positive remaining
constants, with $M(0)=M_0>0$ and $c(0)=W(0)=Y(0)=0$.
Boundary signs preserve $m,c,Z\geq0$.
The bounds $M\leq M_0e^{-\beta t}$,
$W\leq\int_0^tF\,\mathrm{d}t$ and $D\geq1$ give global bounded
solutions, with $m>0$ at every finite time.

Binding cancels in the total repressor balance. Thus
$\rho=F/D$ and $g=v_Ga/F$ give $\dot W=\rho-dZ$ and $V=g\rho$:
Lemma~\ref{thm:storage} applies with $x=W$, $\ell=dZ$ and $h=g$.
There is no explicit branch mismatch; modulation remains through
the RNA pool.

Here dose varies delivered burden RNA; target and regulator RNA
delivery remain fixed. We differentiate with respect to
$\theta=\log k_0$ and write $S_T=\partial_\theta C_T$ in this
section. Here time is in hours, so $C_T$ and $S_T$ have units
$\mathsf C\,\mathrm h$. With $n=M_\theta$, $L=m_\theta$ and $H=-W_\theta$,
the quotient rule gives
\begin{equation}
 \partial_\theta V=v_G\frac{B(\omega n+sL)-ab}{D^2}.
 \label{eq:Vtheta}
\end{equation}
The negative term is direct resource competition. The other term
captures changes in available target RNA, including relief of
repression. Both are evaluated along the resource-coupled trajectory.

\begin{theorem}[Resource-coupled repression certificate]
\label{thm:repression}
For \eqref{eq:repressionbinding}--\eqref{eq:repression} under the stated
assumptions, let nonnegative time-dependent lower and upper bounds on
$m,Z,a,B,b,F,D$ over a dose--loading rectangle have subscripts
$L,U$, with $D_L>0$. Set
$U_U=\epsilon_UF_U/(\kappa_GD_L^2)$ and
$G_U=\max\{d,|\omega U_U-d|\}$, and define
\begingroup
\setlength{\arraycolsep}{3pt}
\begin{equation}
 \mathsf A_U=\begin{pmatrix}
 -\beta-\Delta&\Delta&0\\
 \eta_-+em_U&-\beta-\eta_--e(Z_L+m_L)&em_U\\
 G_U&d+sU_U&-d
 \end{pmatrix}.
 \label{eq:repressioncomparison}
\end{equation}
\endgroup
If $\bar v=(\bar N,\bar L,\bar H)^\top$ starts at zero and solves
$\dot{\bar v}=\mathsf A_U\bar v+(0,0,F_Ub_U/D_L^2)^\top$,
then $|n|\leq\bar N$, $|L|\leq\bar L$, $|H|\leq\bar H$, and
\begin{equation}
 S_T\leq v_G\int_0^T\lambda_T
 \left[\frac{B_U(\omega\bar N+s\bar L)}{D_L^2}
 -\frac{a_Lb_L}{D_U^2}\right]\mathrm{d}t,
 \label{eq:repressionbound}
\end{equation}
where $\lambda_T=(1-e^{-d(T-t)})/d$.
A negative bound proves $S_T<0$ throughout the covered set.
If a finite collection of rectangles covers a whole dose interval
at each prescribed loading, it proves decreasing cumulative output
on that interval.
\end{theorem}

\begin{proof}
Differentiation gives the matrix in \eqref{eq:repressioncomparison}
with actual states and $\omega U-d$ replacing $G_U$, where
$U=\epsilon F/(\kappa_GD^2)$; the forcing is
$(0,0,Fb/D^2)^\top$.
Taking componentwise absolute values bounds the upper right Dini
derivative of $|(n,L,H)|$ by the stated Metzler system, since
$|\omega U-d|\leq G_U$.
Positive comparison gives the sensitivity bounds.
Substitution into \eqref{eq:Vtheta} and integration against
$\lambda_T\geq0$ prove \eqref{eq:repressionbound}.
\end{proof}

The different RNA mechanisms require different sensitivity bounds.
For Model~I, total target RNA is independent of burden, so $n=0$;
the remaining sensitivity system is positive and $L\geq0$.
For Model~II, bound-RNA decay changes the total RNA pool and its
sensitivities need not keep a fixed sign.
The absolute-value comparison accommodates both cases.

Decaying RNA also permits a bound beyond the computed interval.
Set $\zeta=\omega|n|+s|L|$.
Applying \eqref{eq:generaltail} with $\sigma=-1$ gives,
for $T\geq T_0$,
\begin{equation}
 S_T\leq S_{T_0}+\frac{[\partial_\theta Y(T_0)]_+}{d}
 +\frac{v_G}{d}\int_{T_0}^{\infty}\zeta/B\,\mathrm{d}t.
 \label{eq:tailcriterion}
\end{equation}
At $T_1>T_0$, set $M_1=M_0e^{-\beta T_1}$,
$F_1=F(T_1)$ and $b_1\geq b(T_1)$.
Use \eqref{eq:repressioncomparison} with $m_U=M_1$, $Z_L=m_L=0$,
$U_U=F_1/\kappa_G$ and $G_U=d+\omega U_U$ to form
$\mathsf A_1$. If it is Hurwitz and $\bar v_1$ bounds
sensitivity magnitudes at $T_1$, positivity gives
\[
\int_{T_1}^{\infty}\frac{\zeta}{B}\,\mathrm dt
\leq(\omega,s,0)(-\mathsf A_1)^{-1}
\left[\bar v_1+
\begin{pmatrix}0\\0\\F_1b_1/(2\beta)\end{pmatrix}\right].
\]
Finite comparison covers $[T_0,T_1]$. The limiting matrix has
eigenvalues $-\beta,-d,-(\beta+\Delta+\eta_-)$, so a Hurwitz
choice eventually exists. Its adverse budget must still fit
within the accumulated negative margin.

\subsection{Consequences for expression prediction}

We use two parameterizations from one joint fit to our unpublished
48-hour expression measurements. Replicates and channels are
normalized to their own 20 ng anchor; weighted least squares
uses standard errors of the mean (SEM) with a $5\%$ relative
error floor. The archive provides fitted parameters and code
reproducing predictions and certificates \cite{ifflcode};
measurements are not included.

Only burden delivery varies; target and regulator deliveries remain
$M_0=r_0=798.6\,\mathsf C$, where $\mathsf C$ is the model
concentration unit. For $q$ in ng,
$ k_0(q)=K\left(\frac{q}{20\,\mathrm{ng}}\right)^{p_{\rm del}}.$
Here $K/\mathsf C=1781.2128421$ and $p_{\rm del}=1.1605277365$
for Model~I; $K/\mathsf C=1963.9354611$ and
$p_{\rm del}=1.0066005468$ for Model~II.
Both loading configurations retain these maps. Since
$\partial_q C_T=(p_{\rm del}/q)S_T$, nominal-dose and log-burden slope signs agree.
Table~\ref{tab:repression_parameters} summarizes parameters and
status; the repository retains full-precision inputs and initial
conditions.

\begin{table}[H]
\centering
\caption{Repression model parameters (rounded).}
\label{tab:repression_parameters}
\small
\setlength{\tabcolsep}{3pt}
\begin{tabular}{@{}lccc@{}}
\toprule
Symbol & Units & Model~I & Model~II\\
\midrule
$\beta$ & $\mathrm{h}^{-1}$
 & \multicolumn{2}{c}{$0.119389^{\mathrm{F,S}}$}\\
$\Delta$ & $\mathrm{h}^{-1}$
 & $0^{\mathrm{X}}$ & $0.0238778^{\mathrm{D}}$\\
$\omega$ & $1$
 & $0^{\mathrm{X}}$ & $0.0103388^{\mathrm{F}}$\\
$\eta_+$ & $\mathsf C^{-1}\mathrm{h}^{-1}$
 & $0.126387^{\mathrm{F}}$ & $3.89169^{\mathrm{F}}$\\
$\eta_-$ & $\mathrm{h}^{-1}$
 & $0.0669258^{\mathrm{F}}$ & $0.0200000^{\mathrm{F}}$\\
$d$ & $\mathrm{h}^{-1}$
 & \multicolumn{2}{c}{$0.291791^{\mathrm{D,S}}$}\\
$\kappa_K=\kappa_G$ & $\mathsf C$
 & \multicolumn{2}{c}{$51.5856^{\mathrm{D,S}}$}\\
$\kappa_R$ & $\mathsf C$
 & $105.430^{\mathrm{D}}$ & $28.8155^{\mathrm{D}}$\\
$\gamma_RR_{\rm tot}$ & $\mathsf C\,\mathrm{h}^{-1}$
 & $47.8992^{\mathrm{D}}$ & $13.5071^{\mathrm{D}}$\\
$\gamma_GR_{\rm tot}$ & $\mathsf C\,\mathrm{h}^{-1}$
 & \multicolumn{2}{c}{$23.9496^{\mathrm{D,S}}$}\\
\bottomrule
\end{tabular}
\par
{\footnotesize\raggedright
F: directly fitted; D: derived from fitted parameters and fixed
constants; X: mechanism-fixed; S: shared.
$\mathsf C$: uncalibrated model concentration unit.\par}
\end{table}

\begin{figure}[H]
\centering
\includegraphics[width=\columnwidth]{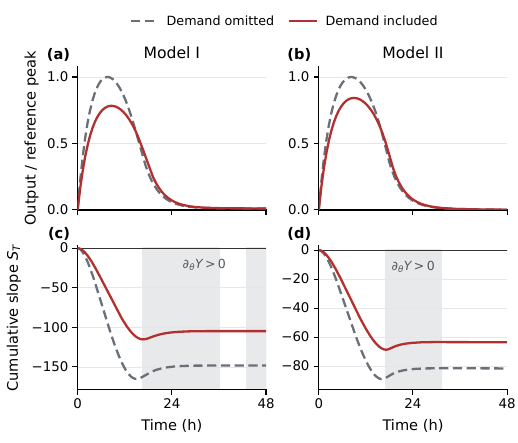}
\caption{Responses at 20 ng. Top: output normalized by the reference
peak ($\epsilon=0$). Bottom: cumulative log-burden sensitivity.
Shading marks positive instantaneous sensitivity.}
\label{fig:repression_response}
\end{figure}

At 20 ng, full target loading changes waveforms by $23.13\%$
for Model~I and $16.65\%$ for Model~II, despite 48-hour endpoint
changes of only $+0.15\%$ and $+2.46\%$
(Fig.~\ref{fig:repression_response}, Table~\ref{tab:repression_metrics}).
RNA decay makes resource demand transient, so these endpoints
miss much larger changes in cumulative output.

\begin{table}[H]
\centering
\caption{Full-loading effects at 20 ng relative to
$\epsilon=0$ (fixed fit).}
\label{tab:repression_metrics}
\small
\begin{tabular}{@{}lrr@{}}
\toprule
Response measure & Model~I & Model~II\\
\midrule
\multicolumn{3}{@{}l}{\textit{Expression changes}}\\
Waveform discrepancy $D_{48}^{\infty}$ & $23.13\%$ & $16.65\%$\\
Peak delay (h) & $1.00$ & $0.69$\\
Endpoint change & $+0.15\%$ & $+2.46\%$\\
Cumulative-output change $E_{48}$ & $-13.35\%$ & $-10.18\%$\\
\addlinespace[2pt]
\multicolumn{3}{@{}l}{\textit{Cumulative dose sensitivity}}\\
Relative slope change $R_{48}$ & $29.33\%$ & $22.13\%$\\
\bottomrule
\end{tabular}
\end{table}

At 20 ng and full loading, both instantaneous sensitivities are
positive at 24 h, while cumulative log-burden sensitivities are
$-106.29$ and $-63.96$. Increasing burden can therefore increase
output at a particular time but reduce its integrated output.

\begin{figure}[H]
\centering
\includegraphics[width=\columnwidth]{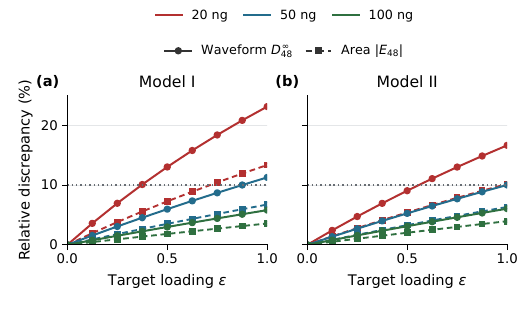}
\caption{Omission errors: circles show $D_{48}^{\infty}$;
squares show $|E_{48}|$. Colors distinguish doses. Lines join
nine sampled loadings without certifying intervening values;
$10\%$ is illustrative.}
\label{fig:omission_tests}
\end{figure}

Quantitative omission decisions remain prediction dependent
(Fig.~\ref{fig:omission_tests}). At full loading and 20 ng,
both mechanisms exceed illustrative $10\%$ waveform and
cumulative-output tolerances.
At 50 ng and full loading, Model~I gives
$D_{48}^{\infty}=11.28314433\%$ and $|E_{48}|=6.74370036\%$;
Model~II gives $10.01999815\%$ and $6.34496388\%$,
respectively. Thus only the waveform tolerance is exceeded in both
models. The Model~II waveform value lies just above the illustrative
cutoff. These comparisons sample three doses and nine loadings;
connecting curves do not certify intervening quantitative errors.

\subsection{Cumulative ordering across doses and horizons}

Validated bounds establish the same cumulative ordering for both
mechanisms, despite their different RNA dynamics.
\begin{corollary}[Cumulative ordering at the fixed fit]
\label{cor:fixedfit}
For each model at the parameterization and delivery map specified
above, $\partial_q C_T(q,\epsilon)<0$ for every
$q\in[20,100]$ ng, $\epsilon\in[0,1]$ and finite $T>0$.
\end{corollary}

\begin{proof}[Validated numerical bounds]
The delivered-burden and loading domains are covered by
$32\times8=256$ rectangles for Model~I and $33\times16=528$
for Model~II. Initially the target-RNA sensitivities vanish, giving
\[
 S_T=-\frac{v_GM_0b(0)}{2D(0)^2}T^2+O(T^3),
 \qquad T\longrightarrow0^+.
\]
Uniform interval bounds establish negative production sensitivity on
the first $0.002$ h for Model~I and $10^{-4}$ h for Model~II.
Propagated bounds then cover every remaining horizon through
$T_0=48$ h. On each rectangle, comparison to $T_1=96$ h and
the analytic tail bound place the adverse future contribution
below the accumulated negative margin. The delivery map has
positive derivative, so nominal-dose and log-burden slope signs agree.
\end{proof}

For Model~I, the uniform bounds are $S_{24}<-57.70$,
$S_{48}<-47.74$ and $S_T<-8.54$ for all $T\geq48$ h.
For Model~II they are $S_{24}<-14.30$,
$S_{48}<-1.91$ and $S_T<-1.68$ for all $T\geq48$ h.
These slopes are with respect to log burden. The sign near zero
is covered separately; no negative constant is claimed uniformly
as $T$ tends to zero.

For Model~II, conservative state enclosures can obscure
the competing sensitivity terms. Higher-degree polynomial approximations
reduce a diagnostic 48-hour upper bound from $+235.97931972$
to $-25.36704987$. The initial cover passes finite-horizon
tests, but its highest-dose interval fails the tail test
($+0.10285109$). Splitting this interval in two gives the accepted
528-rectangle cover, without changing the model or fit.
The code archive provides the grids, validated algorithms,
paired slope bounds and tail budgets, runtimes and replay code.
An inconclusive bound is not evidence of reversal.

\section{Discussion}
\label{sec:discussion}

An omitted coupling should be judged by the prediction it must
preserve. Both repression models retain cumulative dose ordering
despite substantial trajectory and cumulative-output changes.
Preserved direction gives limited information about magnitude
or timing; endpoint agreement can hide transient differences.
Omission decisions should therefore assess ordering and
quantitative accuracy separately, with application-specific
tolerances.

Observation time also matters. Our nonlinear example shows
that stable dynamics permit delayed slope reversal on an
$O(1/\eta)$ timescale under weak pathway mismatch. A finite
simulation can miss later changes in dose dependence. For the
repression models, tail bounds extend validated finite-horizon
calculations to every positive finite observation time over the
stated dose and loading ranges. The observation period must
therefore accompany any ordering claim.

These guarantees concern one fixed fit to endpoint measurements.
Measured time courses are needed to assess the predicted
transients; parameter uncertainty and other architectures
require further analysis. We encourage circuit studies to state
the prediction, reference model, tested domain, observation
period and tolerances, and release model inputs and validation
code. Applying these checks across architectures would clarify
when simpler models support reliable design and when additional
coupling must be retained.


\end{document}